# Review of Mobile Apps Permissions and Associated Intrusive Privacy Threats

**Akosua Boakyewaa Teye, Ezer Osei Yeboah-Boateng**

***Abstract*: The age of technology has created a huge market for smartphones and Apps usage and a new generation has been created based on knowledge sharing. Now knowledge has been made easily accessible by Apps but; are users even aware of the permissions that these Apps require and the privacy issues involved? The study was conducted on the basis of how users make use of Apps. It was conducted through the assessment of permissions required by various Apps through carefully selected third-party Apps and the devices' settings and also a review of existing literature that has been conducted in fields within Apps and privacy. It will be unearthed that a many different but exhaustive lists of permission are sought by each App installed and the device it is installed on can quite give the user the information. Also not all permissions sought were found to be risky but some just created a path or a vulnerable point for other malicious programs to take advantage of.***

## I. INTRODUCTION

The Internet and its attendant applications have become ubiquitous as well as intrusive. Today, we use the Internet and applications for work, for leisure and livelihood. Associated with these interactive transactions are often, unwelcome, intruding, trespassing, yet disruptive threats to the end-users privacy [1].

Have you ever thought of what data your favorite App can access from your smart device? What kind of data do you input on your smart devices? Have you ever pondered on who has access to the data stored on your smart devices?

These and many such questions have been necessitated by the pervasive and ubiquitous technologies which have been exacerbated by the proliferation of Applications (popularly referred to as "Apps". The use of Apps has invariably exploded with the introduction of smart devices and their corresponding mobile Operating Systems, including Google's Android platform and Apple's iOS [2].

Moreover, these Apps create room for major risks to the privacy and subsequently the security of personal and sometimes sensitive information of the user. Sensitive data gathered by these Apps may be made available to third parties without the knowledge of the user [3]. Users are happily predisposed to make their data available because of the convenience of some 'everyday helpful Applications' but most users are oblivious to the amount of personal data being sold on 'black markets'. For the sole convenience of online enjoyments, users are unwary of sharing various forms of personal data ranging from location, personal search, purchasing history and many more.

*"If you have a smart phone or other mobile device, you probably use Apps – to play games, get turn-by-turn directions, access news, books, weather, and more. Apps are easy to download and often free, mobile Apps can be so much fun and so convenient that you might download them without thinking about some key considerations: how they're paid for, what information they may gather from your device, or who gets that information."* [4, p. 1]

The key research questions are:

1. What are the key permissions that the Apps seek on mobile devices?
2. To what extent are the mobile Apps intruding on users' privacy?

This introductory section deals with the background and problems associated with downloading and installing mobile Apps, without and precautions on permission granted to the Apps. It is followed with the Literature review and then the methodology. The results are presented and then followed by some discussions and conclusion.

## II. RELATED WORK

### A. Apps

Mobile communication is so incorporated into our lives that people feel uncomfortable without a cell phone. A while back, the most popular usage of phones were calling and sending texts. However, smart phones (*a multifunctional device that not only communicates, but also helps to learn, earn, and have fun* [5]) give more functions to the user and all these are made possible with the development and onset of mobile Applications (Apps).

According to [5], mobile Applications date back to the end of the twentieth century. But in general, they consisted of small arcade games, ring tone editors, calculators, calendars, and so forth. The onset of the new millennium brought forth a fast market growth of mobile content and Applications. Nokia was prominent for putting the 1970s video game 'Snake' on some of its initial handsets. Others followed, by incorporating games like Pong, Tetris, and Tic-Tac-Toe.

[4] defines a mobile App as *" a software program you can download and access directly using your phone or another mobile device, like a tablet or music player"* [4, p. 1].

Invariably, It can also be explained as consisting of programs designed to make users more productive and/or assist them with personal tasks [6].

The fundamentals for using an App and specifically a mobile app are very simple. The user requires a smart mobile device with internet access. However,

**Akosua Boakyewaa Teye**, West End University College, Kasoa, E-mail: loctovia@gmail.com

**Dr. Ezer Osei Yeboah-Boateng**, Faculty of Computing & Information Systems (FoCIS), Ghana Technology University College, Accra Ghana, E-mail: eyeboah-boateng@gtuc.edu.gh

not all Apps work on every variation of mobile devices. Each device on the market comes with a unique operating system which the user is committed to as long as he uses that particular device which is further bound to a list of pre-loaded Apps that has been made compatible with that device via the "App Store" or "Play Store" of that particular Operating System. All the major mobile operating systems, i.e. Android, iOS, and Windows Mobile among others have incorporated App stores as well as facilitating online presence for users to look for, download, and install Apps.

The idea of App stores as distribution networks for Apps can be traced back to the 1990s but was first used by the Electronic App Wrapper [7] through electronic catalogs that did not necessarily need computers to function. Looking at what an App store does, it can be deduced that an App store is a variation of a marketplace usually digital based that allows for distribution, and purchase of software and can sometimes be on a mobile platform (which would then be known as a mobile App store).

Usually to set up an account with a mobile Apps store, you may have to provide a credit card number, especially if you're going to download an Apps that is not free [4]. There are a wide range of Apps that are available for various purposes including education, health, pleasure, productivity, communication aid in daily activities and others [6]. Some are as follows:

i. Play stores or App stores: Examples include Google Play store, 1 Mobile Market, Samsung Market, Microsoft Apps, Apple App store and others. These are the pioneer market places where all other Apps can be purchased and/or downloaded from.
ii. WhatsApp: A free messaging App for Android and other smart phones. Enables individual and group messaging, voice and video calls and other functionalities [8] [9]
iii. Calendar: An average App with any smart device. A user can use this App like any other calendar; plan (Add in your classes, test dates, homework time, work, Appointments). If used efficiently the calendar makes sure that a user never forgets anything. It is technically a walking reminder for events and can be synchronized to emails sometimes.
iv. Messenger: An App that is similar to texting or used chatting. The advantage is that it is usually free so the user does not have to pay for every message like texting using short messaging service (SMS).
v. Instagram : An App that can be used for photo and video sharing and it's free. [10]
vi. Cloud Storage Apps: Enables Cloud storage (for pictures, videos, documents, and many other files) between devices. An example is Dropbox which allows storage of up to 5GB (usually free), saves space on phones, computers, and tablets. [11]. Others include OneDrive from Microsoft, Google Drive (A free App that can help create 15 GB cloud that can store various types of files without taking up space on any of your devices. Information can be accessed wirelessly, and all information can be organized into customizable folders. All you need is a Google account [12]), Amazon Cloud Drive and Box [13]. All the cloud Apps have similar functionalities.
vii. Pages: $9.99 iOS App. It is easy to use for typing, formatting, editing, and organizing papers and documents. Different formats, such as landscape, research paper, essays, letters, resumes and newsletters.
viii. iDraw: it is worth $2.99 App. It includes hundreds of drawing tools, formatting options, and colors to create a variety of things. Different papers, such as graph paper, yellow lined paper, and clear white. The user is able to import photos, create structural layouts, and clean up any image with an assortment of filters.
ix. YouTube: YouTube is not just funny cat videos. YouTube can help with classes ranging from algebra and chemistry, how to correctly add citations to papers, lectures on history, and much more. It is a free App.

For the purposes of this study, Apps represent all manner of mobile device Applications, be they pre-installed, or customized, or user installed, whether paid for or free. The Apps aid users with everyday tasks and in turn "hook" some users; i.e. "attached" or "addicted" to certain Apps, to the extent that they cannot fathom a day without some of these Apps; a form of technology addiction [6]. Consequently, some users, in spite of the possible dire consequences, may infringe on their own privacy in using the Apps and bear the affects there from.

*1) Privacy*

*Stewart et al*, [14] assert that privacy can be a difficult entity to define. They however stated in their book Certified Information System Security Professional Study Guide, that the term is used often in many contexts without much quantification or qualification. They explained Privacy as "*Active prevention of unauthorized access to information that is personally identifiable (that is, data points that can be linked directly to a person or organization). Also as Freedom from unauthorized access to information deemed personal or confidential or Freedom from being observed, monitored, or examined without consent or knowledge"*

Privacy can also be explained as an indispensable human right that includes the right to be left alone [15] as cited by [16]. Simply stated, Privacy is a human right that enables one to have control over what other people are privy to about oneself. It also means that there should be some level of trust between two individuals for them to let each other know their private issues.

This study adapts the privacy dimension as safeguarding any violations or compromises resulting from unauthorized collection, access and usage of personable or corporate data, whether in transit or at rest.

When users allow Apps to access some aspect of their phones, they trust the Apps and subsequently the developer [17]. It is therefore paramount to let users know what data are collected and used for, as well as the Security policies employed through the end-user license agreement (EULA).

*2) Using Apps*

Apps can access most data from users but the user need to give out access or permit the App to access certain data before it can. However,

Most Apps today are structured in a form that does not allow the user to use and benefit from it if the user does not allow access to certain data.

When a user registers with an App store or even download Apps in seclusion, he may be asked for permission to let the Apps access information from his device. Some Apps may be able to access: the user's device (phone, tablet, laptop, or any other device), contacts (phone and email contacts), calendar and internet data, call logs, device's location, camera or photos, and sometimes even the device's unique IDs. [18] [19] [20]

However, information about how the App is used itself may also be made vulnerable and accessed. Notwithstanding all these, some Apps access only the data needed to carry out what it needs to do whist others even access data that's not related to the core function as well as other functions of the App.

> *"If you're providing information when you're using the device, someone may be collecting it – whether it's the App developer, the App store, an advertiser, or an ad network. And if they're collecting your data, they may share it with other companies".* [4, p. 1]

So the question remains: *how can a user know the type of data that an App will access or even share?* It's not entirely simple for a user to have knowledge on what information a particular App will access, and even if the user somehow did, he may not know how it will be used.

*3) WHAT TO DO…*

As a prospective or an existing App user, before you consider an App:

i. Take into consideration your knowledge [5] about the App's creators and its function as well. (The App store you decide to download from may take account of the information about the company that developed the App; that is if the developer makes that information available. However, if the developer doesn't give out contact information –mostly like a website or email address – the App may be considered less than trustworthy.) In other words scrutinize each app well enough before you decide to install it especially finance-related apps [6]

ii. After an App has been installed, one can access the permissions allowed for the app in the devices settings.

iii. Popular operating systems like the Android operating system and the Google Play Store, gives the user an opportunity to read the "App permissions" just before the App is installed.

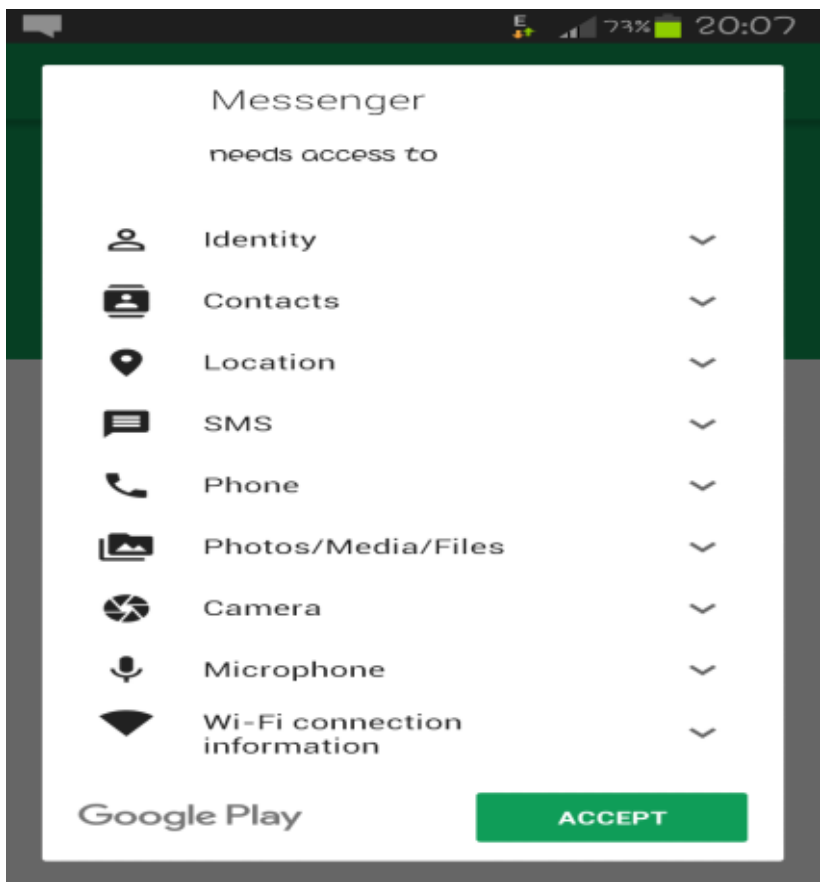

**Figure 1a: Sample Apps Permissions**

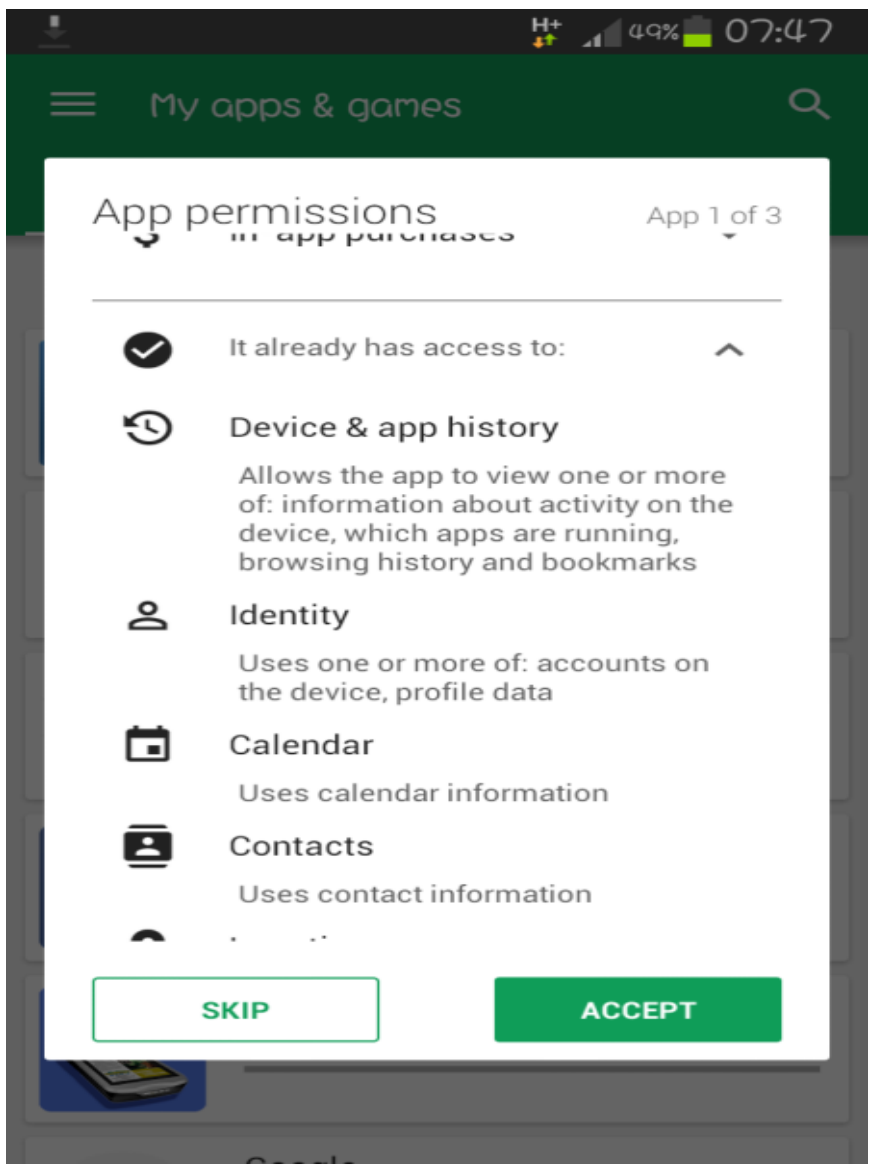

**Figure 1b: Sample Apps Permissions**

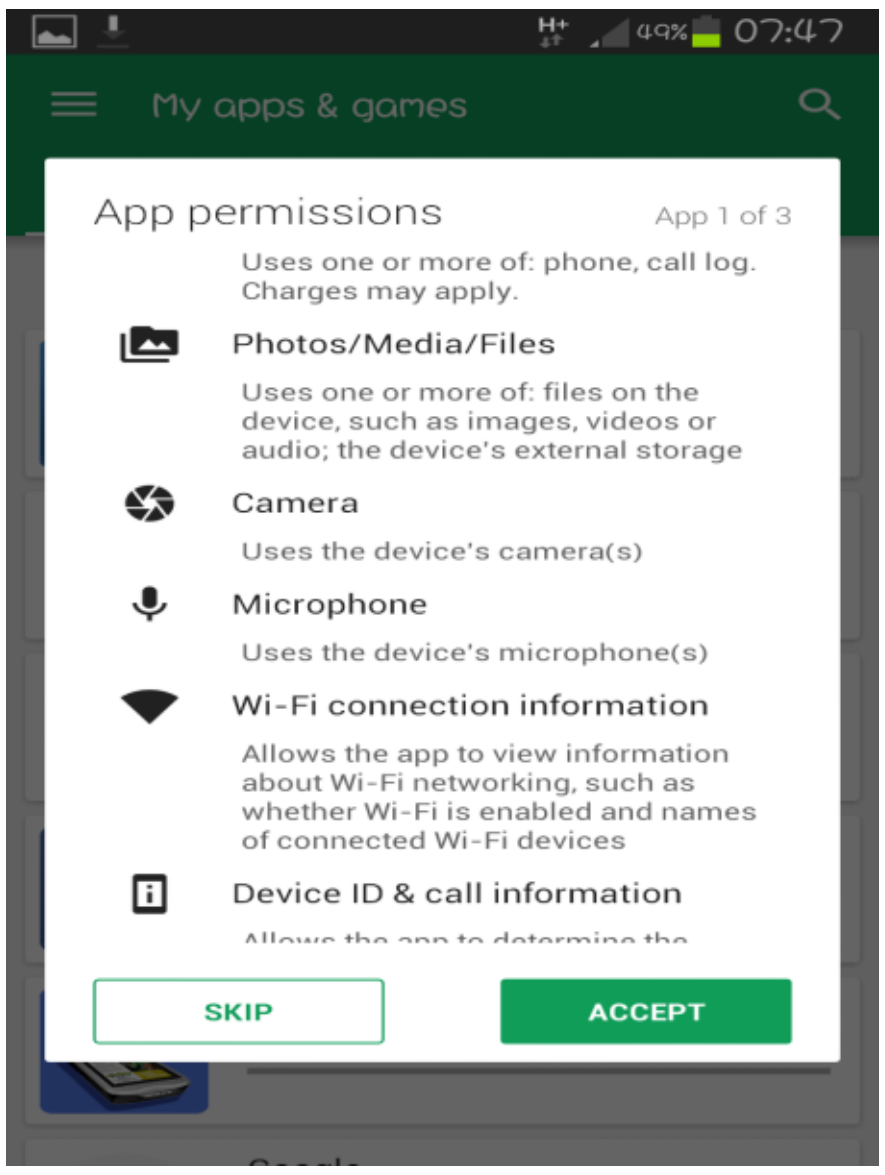

**Figure 1c: Sample Apps Permissions**

These data are there for the benefit of the user. So the user is advised to educate himself by taking the time to read. Studies by [18] and [19] have shown that users are woefully under-informed and educated on App permissions. It is valuable and handy information that informs you about the App and what it will access from your device. However it should be analyzed if the information add up given the core function of the App; for example, there's no reason for a "kids alphabet game" or "3D picture" App to read your text messages.

Despite all the above, it is still difficult for users to adhere to some of the cautions stipulated because of the opt-in policies (*to give financial institutions the right to share personal data, such as annual earnings, net worth, employers, personal investment information, loan amounts, and Social Security numbers, to other institutions.* [21]). Figures -1a, -1b and -1c above all show that the user is given only two options for any given Apps installation or upgrade i.e. SKIP or ACCEPT. If the user decides that the Apps is taking too much information that it may not need, the only unfortunate outlet is not installing the App or uninstalling it if the Apps has already been installed.

*1) Data Access By Apps*

Some Apps utilize precise location data to give you maps, routes to emergency service locations, restaurants, residences, or even route you through an unknown territory. Some give out location data to ad networks, which they use in target marketing usually based on combined information in their databases about what they may already have on the user. However, once an App has your permission to access your data (whether contacts, photos, location and others); it can do so until the user changes the settings on his device. Turning off location and GPS in your phone settings can to some extent curb off the sharing of your location with the Apps on your device but this will in turn hinder some of the Apps especially those that rely on location data to work efficiently.

## III. METHODOLOGY

This is an exploratory study on the perceived intrusive threats posed as a result of mobile Apps permissions granted and/or escalated privileges on mobile devices. First, targeted mobile devices samples were profiled to ensure that they were installed from genuine Apps developer's sites. This was necessary to ensure that hackers-created or cloned fake Apps versions were ruled out. Also, the test devices were checked and ascertained as secured; verifying that the devices would disallow unauthorized access and installations by any Apps, bypassing operating systems security (Sthanu, 2015).

Three (3) mobile Apps profiler tools were purposively selected for the use of profiling the various Apps permissions. These are Apps Detective, Suspicious Apps and Permission Friendly Apps.

Within the context of this research, App profilers are Apps that can assess (or evaluate) and access the extent of security and vulnerability as well as permissions sought by other Apps on any given device. In summary, they are Apps that can track the performance of other Apps.

*a) Apps Detective*

An Apps designed for phones by zmarties. It takes a closer look at the finer details inside the apps installed on any given device. It does not require any permission to function.

*b) Suspicious Apps*

An Apps created by D. Schwarz that aids in finding other Apps on any device that requests for privileges that are considered suspicious and rates the Apps based on that. A higher percentage shows that the said App requests for many 'suspicious permissions' and a lower percentage deems the App as relatively safe in the area of permission-seeking and data access.

*c) Permission Friendly Apps*

Developed by [22], this App is considered a companion for detecting applications that require potentially abusive permissions. It aids in checking the Apps for risky permissions. It indicated whether a permission being solicited was harmless (with a heart symbol but indicated in the abridged table of this research in the colour-green) or considered risky (with a caution sign- indicated with red in the abridged table).

The datasets derived from using the three profilers and the default settings of the devices used were then text mined with [23]. According to [24], Text mining tools are readily available to aid businesses as well as individuals and researchers analyze data. They are able to extract key elements from large unstructured data sets, discover patterns and relationships, and summarize the information. This was employed to create taxonomy of key permissions and accessible features.

## IV. RESULTS AND DISCUSSION

The study was initiated by gathering data which included using the three profilers mentioned as well as the 'device default settings' to access the permissions that any installed App requests.

Each installed App was subjected to the same process. Device default permissions of selected Apps were first evaluated. This was done by checking the App permissions requested under the device settings. Then the same App was then evaluated with the Profilers one after the other for the permissions that has been requested, and analyzed or recorded and unearthed by the profilers.

All these were tabulated but due to the enormous data sets, the table has been abridged. The table is inserted in the index The data included some of the popular apps like Messenger, WhatsApp, Snapchat, VLC and others for inclusion in the table. The apps that was used in this datasets was categorized broadly under Games, Social networks, health and lifestyle apps, Profilers, Browsers, Education apps and Entertainment apps. The entire dataset was text-mined using [23] as explained earlier to highlight the key items or variables that had prominence in the datasets.

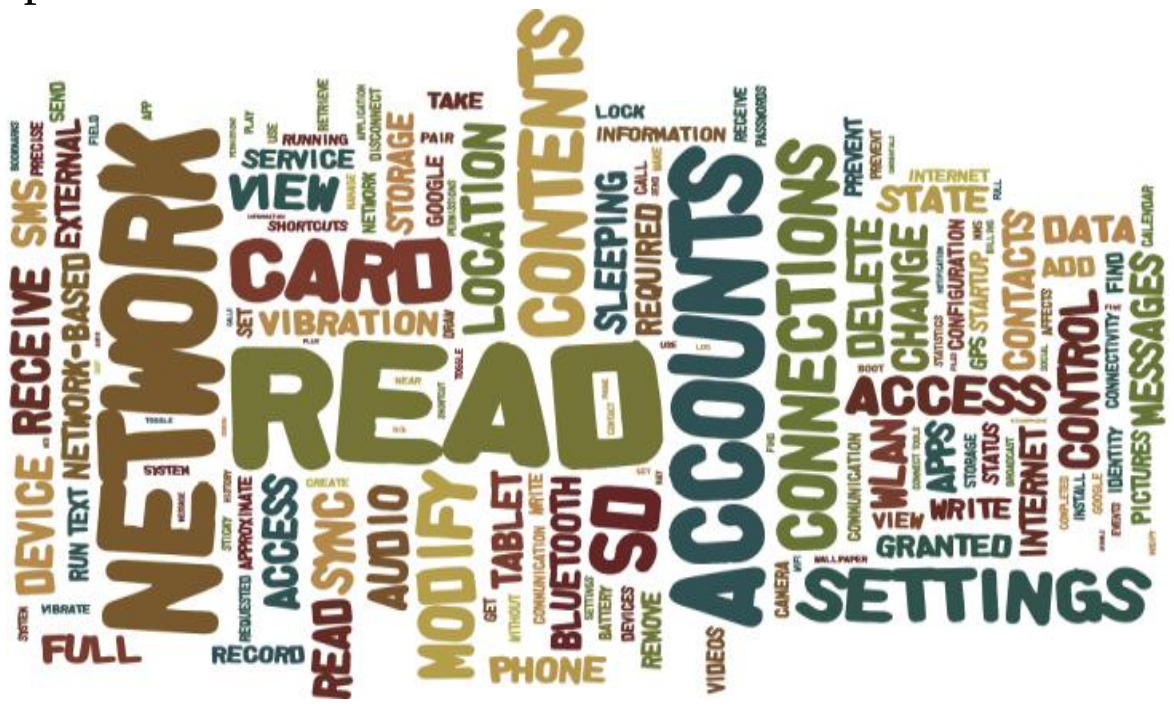

**Figure 2: Taxonomy of Mobile Apps Permissions (text-mined using Wordle.com)**

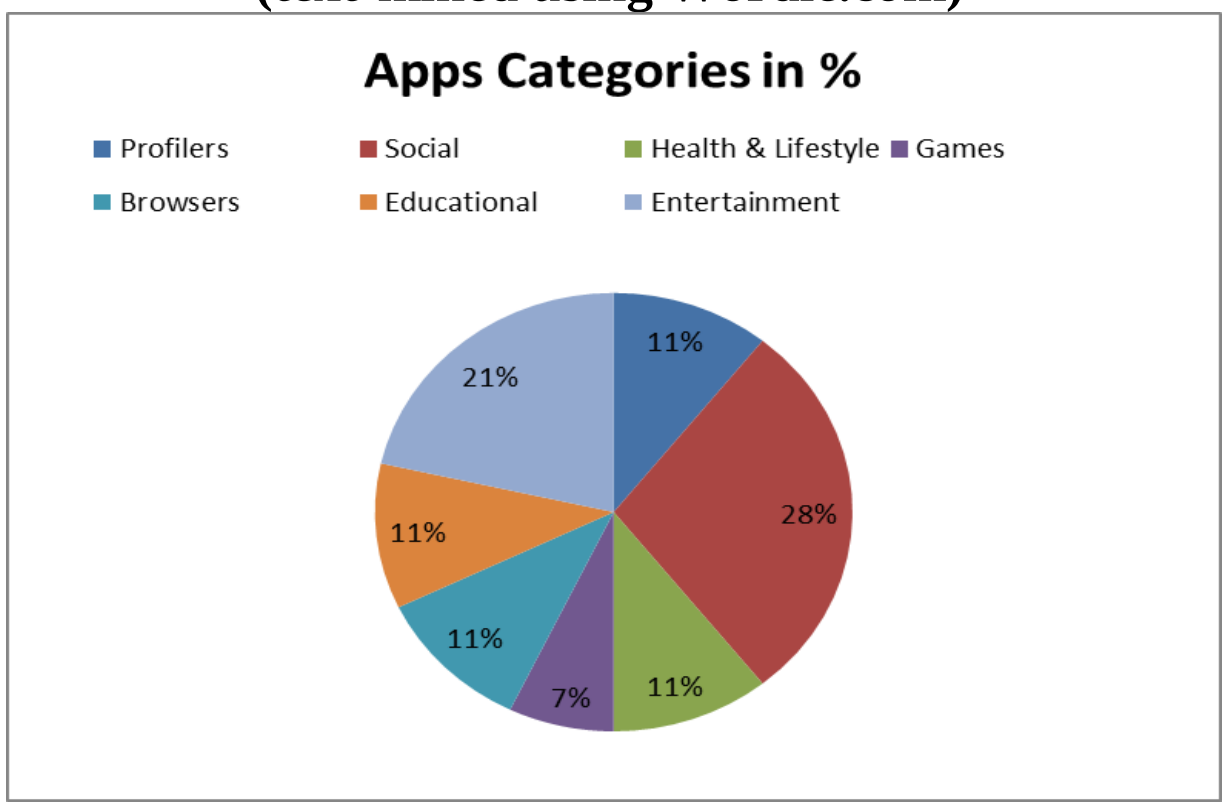

**Figure 3: Apps Categories in Percentages**

The data gathered showed that a majority of the apps were Social Network and Entertainment Apps with a total percentage of 28% and 21% respectively. This signifies the proliferation of social terrain and the extent to which lives are impacted with them.

Some key observations made especially with respect to the Social network Apps were:

- From the analysis made with Suspicious Apps, some social network Apps could receive a rating of as high as 86% with the lowest rating being 14% while the profilers themselves could obtain percentage rating of as low as 0% and as high as 1%. Also for the other categories of app, the percentage rating hovered between 2% and 20%.
- App Detective's analysis also affirmed that the Social Apps were soliciting for more permission that the other types of Apps. From that platform, it was evident that the Profilers sought for no permissions to as high as two (2) permissions to function. On the other hand, the social Apps sought for as low as 17 requested permissions on a device to as high as 57 individual permissions to function effectively.

From the text-mined data, it was discovered that the most prominent of permissions sought by Apps are the:

i. **Read:** this includes Apps reading the contents of the storage medium that is available in the device, text messages, battery statistics, phone identity, calendar events and confidential data, google service configuration and others.

Some of these 'read' permissions may seem harmless enough for the device that it's being used on but the Profilers (Permission Friendly App profiler) categorized all the permissions under the harmfulness or harmlessness of the said permission. Examples of permissions that may harm the user is the permission that

- Allows the app to read the device's status and identity. This permission *allows app to determine the phone number and the device ID, whether the call is active and the remote number connected by a call)*
- Allows the app to read text messages. This permission may cost the user some monetary charges and also opens another doorway for -*malicious apps to take advantage of*.
- Allows the app to read calendar events plus *confidential data*

Some permissions are however not detrimental to the user and some of them may be Read Google Service configuration and Reading the contents of the SD card.

ii. **Network:** Most Apps depend on a network to function effectively so the network being prominent among the text mined data was not unexpected. Asking to access full network Access and to connect and disconnect from WLAN are two network permissions that was deemed malicious by the Permission Friendly App. Some other permissions sought under network that was deemed ok are listed below:
    - View WLAN connections
    - View network connections
    - Change network connectivity
    - Control near field communication,
    - receive data from the internet

iii. **Accounts**
    - Add or remove accounts
    - Create accounts and set passwords
    - Find accounts on the device
    - Get and manage accounts

iv. **Modify Settings**
    - Modify or delete contents of device's SD card
    - Modify device's contacts
    - Modify system settings

v. **Connections**: These sets of permissions were related to the network permissions

vi. **Contents**: these sets of permissions were related to the read permissions

vii. **SD card:**
    - Modify or delete the contents of your SD card
    - Read the contents of your SD card

After critically studying the data, it was unearthed that generally, most apps initiate at startup and request for permissions that require the screen of your device to stay active. This invariably will not tamper with any data but will drain the devices' battery faster. Other permissions that are solicited by apps included but were not limited to recoding audio by using device's microphone, taking pictures and videos, controlling the vibration feature of the device and access course and fine location using GPS and network.

Some of the permissions were executed with the knowledge of the user and others were like a 'ripple effect'. This means one permission granted that will create a vulnerability hole for other Apps and malicious codes to take advantage of. The Receive text messages permission could allow an app to monitor or delete messages sent to a device without alerting the user of that device.

According to the Profilers (Suspicious App) the social Apps which constitute 28% of the total datasets solicit for more permissions that the other types of apps. This feature was found to be unpleasant because these social apps had a high percentage which was a feature that preached danger.

The three chosen Profilers were used to assess each other, it was noticed that Permission Friendly Apps and App Detective did not require any special permission to function effectively and therefore did not seek any permissions at all from the devices they resided on but 'Suspicious App' which sought for permission only did so minimally.

## V. CONCLUSION AND FUTURE SCOPE

Many users of Apps are uninformed about privacy issues that relate to the Apps they use daily even though privacy mechanisms are getting more refined as well as complicated with time It was unearthed that even though Apps request for a variety of permissions from users, not all of them are harmful to that user, some Apps may just create room for malicious codes to take advantage of. Also asking for more permission does not guarantee that the App is safe enough to use on the contrary.

This study has taken a further step towards the bridge between humans, smart devices and information privacy to understand why certain decisions may be taken in helpful Apps usage. There is however still more work that can be done in this area.

## VI. CONCLUSION

A conclusion section is not required. Although a conclusion may review the main points of the paper, do not replicate the abstract as the conclusion. A conclusion might elaborate on the importance of the work or suggest applications and extensions.

## REFERENCES

1. E. O. Yeboah-Boateng, Cyber-Security Challenges with SMEs in Developing Economies: Issues of Confidentiality, Integrity & Availablity (CIA), Copengahen: center for Communications, Media & Information technologies (CMI), Aalborg University, 2013a, pp. 1-217.
2. D. Schulke, The Regulatory Arms Race: Mobile-Health Applications and Agency Posturing, 2013.
3. R. Adhikari and D. Richards, "Security and Privacy Issues Related to the Use of Mobile Health Apps," Auckland, New Zealand, 2012.
4. Federal Trade Commission , "Understanding Mobile Apps," 2011. [Online]. Available:https://www.consumer.ftc.gov/articles/0018-understanding-mobile-apps. [Accessed 9th December 2016].
5. J. F. Clark, History of Mobile Applications; Theory and Practice of Mobile Applications, 2012.
6. M. E. Vermaat, S. L. Sebok, S. M. Freund and J. T. F. M. Campbell, Discovering Computers 2016; Tools, Apps, Devices and the Impact of Technology, Boston, USA: Cenage Learning, 2016.
7. K. Ford, "Ellectronic AppWrapper," 2008. [Online]. Available: http://www.Kevra.org/TheBestOfNext/SWCatalogs/page274/page274.html. [Accessed 27th January 2017].
8. W. Inc, "Whatsapp," 2016. [Online]. Available: http://www.whatsapp.com. [Accessed 27 January 2017].
9. "GDP Tech Tips Sheets WhatsApp," [Online]. Available: http://www.fosi.org/documents/71/Gdp_Tech_Tips_Sheets_WhatsApp.pdf. [Accessed 27 January 2017].
10. Instagram, "Instagram," 2016. [Online]. Available: https://www.instagram.com/?hl=en. [Accessed 26 November 2016].
11. Dropbox, "Dropbox," 2016. [Online]. Available: www.dropbox.com. [Accessed 13 December 2016].
12. "Google Drive," 2016. [Online]. Available: https://www.google.com/drive. [Accessed 28 December 2016].
13. S. Mitrooff, "OneDrive,Dropbox, google drive and Box: Which cloud is right for you?," 2016. [Online]. Available: https://www.cnet.com/how-to/onedrive-dropbox-google-drive-and-box-which-cloud-storage-is-right-for-you/. [Accessed 27 January 2017].
14. J. Stewart, E. Tittel and M. Chapple, CISSP: Certified Information System Seccurity Professional Study Guide, 4th Edition ed., Indianapolis, Indiana: Wiley Publishing, Inc, 2008.
15. P. Siani and B. Azzedine, "Privacy, Security and Trust Issues Arising from Cloud Computing," Bristol, UK, 2010.
16. E. O. Yeboah-Boateng and A. B. Appiah-Nketia, "Multi-Tenancy Issues with Service Delivery in Developing Economies: Privacy, Trust and Availability Concerns," International Journal of Computer Application (2250-1797), vol. 5, no. 6, pp. 82-94, 2015.
17. J. Larimer and K. Root, Security and Privacy in Android Apps, Gogle Developers, 2012.
18. J. Golbeck and M. L. Mauriello, "User Perception of Facebook App Data Access: A Comparison of Methods and Privacy Concerns," Future Internet, 2016.
19. K. Strater and H. Lipford, "Strategies and struggles with privacy in an online social networking community," Liverpool, UK,, 2008.
20. Debatin, J. Lovejoy, A. Horn and B. Hughes, "Facebook and online privacy: Attitudes, behaviors, and unintended consequences," p. 83–108, 2009.
21. G. W. Reynolds, Ethics in Information Technology, 3rd Edition ed., USA, 2010.
22. androidsoft.org, "Permission Friendly Apps - Version 1.4.1," 2014. [Online]. Available: http://54.154.155.50/permission-friendly-apps.html. [Accessed 10 March 2017].
23. Wordle, "Wordle," 2017. [Online]. Available: www.wordle.net/create. [Accessed 16 March 2017].
24. K. C. Laudon and J. P. Laudon, Management Information Systems, Managing the Digital Firm, 12th ed., New Jersey: Pearson Prentice Hall, 2012.
25. Apps, 2016.

**Akosua Boakyewaa Teye** is a Lecturer in Computer Science at the West End University College, Kasoa, in Ghana. She holds a Masters degree in Management Information Systems (MIS) from Coventry University.

**Dr. Ezer Osei Yeboah-Boateng** is the Dean, Faculty of Computing & Information Systems (FoCIS), at the Ghana Technology University College (GTUC), in Accra. Ezer is a Telecoms Engineer and an ICT Specialist; an executive with over 20 years of corporate experience and about 9 years in academia. He has over 20 peer-reviewed international journal papers to his credit. His research focuses on cyber-security vulnerabilities, digital forensics, cyber-crime, cloud computing, Big data and fuzzy systems.

| | | | | | |
|---|---|---|---|---|---|
| | | **APP PERMISSIONS** | | | |
| | | **PROFILERS** | | | |
| | **APP** | **Suspicious App analysis** | **Permission Friendly Apps** | **App detective Analysis** | **Device Settings** |
| 1 | Ali Express | 3%<br>• View WLAN connections<br>• Prevent tablet from sleeping<br>• Retrieve running Apps<br>• Full network access<br>• Modify or delete contents of your SD card<br>• View network connections<br>• Delete other Apps' caches<br>• Read phone status and identity<br>• Modify system settings (*malicious Apps may corrupt your system's configuration*)<br>• Run at startup<br>• Take pictures and videos<br>• Control vibration<br>• Change system display settings<br>• Change network connectivity<br>• Connect and disconnect from WLAN<br>• Read the contents of your SD card | Network Communication<br>• View WLAN connections<br>• Full network access<br>• View network connections<br>• Change network connectivity<br>• Connect and disconnect from WLAN<br>Affects Battery<br>• Prevents Tablet from Sleeping<br>• Control Vibration<br>Your Application Information<br>• Retrieve running Apps<br>• Run at startup<br>Storage<br>• Modify or delete contents of your SD card<br>• Read the contents of your SD card<br>Phone Calls | 17 Requested Permissions<br>• Access wifi state<br>• Wake lock<br>• Get tasks<br>• Internet<br>• Write external storage<br>• Access network state<br>• Delete cache files (REQUIRED BUT NOT GRANTED)<br>• Read settings (REQUIRED BUT NOT GRANTED)<br>• Read phone state<br>• Write settings<br>• Receive boot completed<br>• Camera | • Read phone status and identity<br>• Take pictures and videos<br>• Modify or delete the contents of your SD card/read the contents of your SD card<br>• Change network connectivity, connect and disconnect from WLAN, full network access, view network connections, view WLAN connections |
| | | | • Read the phone status and identity (*allows App to determine the phone number and the device ID, whether the call is active and the remote number connected by a call*)<br>System Tools<br>• Modify system settings<br>Camera<br>• Take pictures and videos<br>Development tools<br>• Change and display settings | • Vibrate<br>• Change configuration (REQUIRED BUT NOT GRANTED)<br>• Change network state<br>• Change wifi state<br>• Read external storage | • Retrieve running Apps, Run at startup<br>• Control vibration, prevent tablet from sleeping<br>• Modify system setting |
| 2 | App Detective | 0%<br>This App does not request any rights | This App requires no permission, no adds and its totally secure | N/A | • No default settings |
| 3 | Suspicious Apps | 1%<br>• Full network access<br>• View network connections | Network Communication<br>• Full network access<br>• View network connections | 2 requested Permissions<br>• Internet<br>• Access network state | • Full network access<br>• View network connectio ns |
| 4 | Permission Friendly Apps | | This App requires no permission, no adds and its totally secure | N/A | • No defaults set |
| 5 | Assassin's Creed III | 7%<br>• Full network access<br>• View network connections<br>• View WLAN connections<br>• Read phone status and identity<br>• Prevent tablet from sleeping<br>• Modify or delete contents of your SD card<br>• Read the contents of your SD card<br>• Send SMS messages (*may cost you unexpected charges and malicious Apps may cost you money by sending messages without your confirmation*)<br>• Receive text messages (*App could monitor or delete messages sent to your device without showing them to you*)<br>• Read your text messages (SMS or MMS)<br>• Connect or disconnect you from WLAN<br>• Change network connectivity<br>• Control vibration | Network Communication<br>• Full network access<br>• View network connections<br>• View WLAN connections<br>• Connect and disconnect from WLAN<br>• Change network connectivity<br>Phone Calls<br>• Read phone status and identity<br>Affects Battery<br>• Prevents Tablet from Sleeping<br>• Control Vibration<br>Storage<br>• Modify or delete contents of your SD card<br>• Read the contents of your SD card<br>Your messages<br>• Send SMS messages<br>• Receive text messages (SMS)<br>• Read your text messages (SMS or MMS | 13 Requested Permissions<br>• Internet<br>• Access network state<br>• Access wifi state<br>• Read phone state<br>• Wake lock<br>• Write and read external storage<br>• Send, receive and read SMS<br>• Change WIFI state<br>• Change network state<br>• vibrate | • Read phone status and identity<br>• Read your text messages (SMS or MMS), receive text messages (SMS), send SMS messages (this may cost you money-malicious Apps may take advantage of this)<br>• Modify or delete contents of your SD card, read contents of your SD card<br>• Change network connectivity, connect and disconnect from WLAN, full network access, view network connections, view WLAN connections<br>• Control Vibration, prevent tablet from sleeping |
| 6 | BBM | 29%<br>• Full network access<br>• Control vibration<br>• Take pictures and videos<br>• Control near field communication | Network communication<br>• Full network access<br>• Control near field communication<br>• View network connections | 32 Requested Permissions<br>• Internet<br>• Vibrate<br>• Camera<br>• Nfc | • Read phone status and identity<br>• Send SMS messages (this may cost you |
| | | • Read phone status and identity<br>• View network connections<br>• View WLAN connections<br>• Read sensitive log data<br>• Retrieve running Apps<br>• Modify and delete contents of your SD card<br>• Read your contacts<br>• Send SMS messages<br>• Record audio (*can record anytime without your permission*)<br>• Set wallpaper<br>• Modify your contacts<br>• Precise location (GPS and network -based)<br>• Add or remove accounts<br>• Create accounts and set passwords<br>• Find accounts on the device<br>• Read calendar events plus confidential information<br>• Add or modify calendar events and send email to guests without owners' knowledge<br>• Read contents of your SD card<br>• Change your audio settings<br>• Pair with Bluetooth devices<br>• Send sticky broadcast (*excessive use can cause the device to be slow or unstable by using too much memory*)<br>• Approximate location (network based)<br>• Prevent tablet from sleeping<br>• Receive data from the internet<br>• Google play Billing service<br>• Read google survive configuration<br>• Run at startup (*can cause a longer booting process and slow down the tablet by always running*) | • View WLAN connections<br>• Receive data from the internet<br>• Google Play billing Service<br>Affects Battery<br>• Control vibration<br>• Prevent tablet from sleeping<br>Camera<br>• Take pictures and videos<br>Phone calls<br>• Read the phone status and identity<br>Development Tools<br>• Read sensitive log data<br>Your Application Information<br>• Retrieve running Apps<br>• Run at startup<br>Storage<br>• Modify or delete the contents of your SD card<br>• Read the contents of your SD card<br>Your Social Information<br>• Read your contacts<br>• Modify your contacts<br>Your messages<br>• Send SMS messages<br>Microphone<br>• Record audio<br>Wallpaper<br>• Set wallpaper<br>Your location<br>• Precise location (GPS and network-based)<br>• Approximate location (network-based)<br>Your accounts<br>• Add or remove accounts<br>• Create accounts and set passwords<br>• Find accounts on the device<br>• Read Google Service configuration<br>Your Personal Information<br>• Read calendar events plus confidential data<br>• Add or modify calendar events and send email to guests without owners' knowledge<br>Audio Settings<br>• Change your audio settings<br>Bluetooth<br>• Pair with Bluetooth devices<br>System Tools<br>• Send sticky broadcast | • Read contacts and phone state<br>• Access network and WIFI state<br>• Read logs (REQUIRED BUT NOT GRANTED)<br>• Get tasks<br>• Write external storage<br>• Send sms<br>• Record audio, set wallpaper<br>• Write contacts<br>• Access file location<br>• Get, Manage and authenticate accounts<br>• Read and write calendar<br>• Read external storage<br>• Modify audio settings<br>• Bluetooth,<br>• wake lock,<br>• broadcast sticky<br>• Receive boot completed | money)<br>• Take pictures and videos<br>• Record audio<br>• Approximate location (network based), precise location (GPS and network based)<br>• Modify contacts and read your contacts<br>• Add or remove accounts, create accounts and set passwords, find accounts on the device, read google service configuration<br>• Control near field communication, full network access, Google Play Billing, receive data from the internet, view network connections, view WLAn connections<br>• Pair with Bluetooth devices<br>• Retrieve running Apps, run at startup<br>• Control vibration, prevent tablet from sleeping<br>• Set wallpaper<br>• Change your audio settings<br>• Send sticky broadcast |
| 7 | Black Player | 2%<br>• Prevent tablet from sleeping<br>• Change your audio settings<br>• Full network access<br>• View network connections<br>• Modify or delete contents of your SD card<br>• Read phone status and identity<br>• Pair with Bluetooth devices<br>• Control vibration<br>• Run at startup<br>• Read the contents of your SD card | Affects Battery<br>• Control vibration<br>• Prevent tablet from sleeping<br>Audio Settings<br>• Change your audio settings<br>Network Communication<br>• Full network access<br>• View network connections<br>Storage<br>• Modify and delete the contents of your SD card<br>• Read the contents of your SD card<br>Phone Calls<br>• Read the phone status and identity<br>Bluetooth<br>• Pair with Bluetooth devices<br>Your Application Information<br>• Run at Startup | 10 requested Permissions<br>• Wake lock<br>• Modify audio settings<br>• Internet<br>• Access network state<br>• Write external storage<br>• Read phone state<br>• Bluetooth<br>• Vibrate<br>• Receive boot completed<br>• Read external storage | • Read phone status and identity<br>• Modify and delete contents of your SD cars, read the contents of your SD card<br>• Full network access, view network connections<br>• Pair with Bluetooth devices<br>• Run at startup<br>• Control vibration, prevent tablet from sleeping<br>• Change your audio settings |
| 8 | Browser | N/A | N/A | 28 requested Permissions<br>• Access and install drm<br>• Access course location<br>• Access download manager (REQUIRED BUT NOT GRANTED)<br>• Access fine location, access network state, access wifi state<br>• Get accounts<br>• Use credentials<br>• Internet<br>• Nfc<br>• Set download completed intents (REQUIRED BUT NOT GRANTED | • Directly call phone numbers (this may cost you money)<br>• Approximate location (network-based), precise location (GPS and network-based)<br>• Read your contacts<br>• Read your own SD card<br>• Read your web bookmarks and history, write web bookmarks and history<br>• Modify or delete contents of your SD card, |
| | | | | )<br>• Set wallpaper, wake lock<br>• Write external storage, write settings<br>• Read and write sync settings<br>• Manage accounts<br>• Read profile, read contacts, call phone<br>• Write source settings<br>• (REQUIRED BUT NOT GRANTED)<br>• .permission.read history bookmarks<br>• .permission.write istory bookmarks<br>• Com.android.launcher.permission.install shortcut<br>• Change network state,<br>• read external storage | read the contents of your SD card<br>• Add or remove accounts, find accounts on the device, use accounts on the device<br>• Change network connectivity, control near field communication, full network access, view network connections, view WLAN connections<br>• Prevent tablet from Sleeping<br>• Set wallpaper<br>• Read sync setting, toggle sync on and off<br>• Install shortcuts, modify system settings (*malicious Apps may corrupt your system's configuration*)<br>• Access DRM content, Install DRM content |
| 9 | Calculator | N/A | N/A | N/A | • No default settings |
| 10 | Calendar | N/A | N/A | 15 Requested permissions<br>• Write external storage<br>• Get, manage accounts<br>• Internet<br>• Vibrate<br>• Read contacts, calendar<br>• Write calendar<br>• Wake lock<br>• Use credentials<br>• Read sync settings<br>• Receive boot completed<br>• Read sms<br>• Com.googl e.android. googleApps .permission .google. auth.mail<br>• Read external storage-+<br>• | • Read your text messages( SMS or MMS)<br>• Read your contacts<br>• Add or modify calendar events and send email to guests without owners' knowledge, read calendar events plus confidential information<br>• Modify or delete contents of your SD card, read the contents of your SD card<br>• Add or remove accounts, find accounts on the dev ice, Google mail, use accounts on the device<br>• Full network access<br>• Run at startup<br>• Control vibration, prevent tablet from sleeping<br>• Read sync settings |